# A critical assessment of the dislocation-driven model for superplasticity in yttria tetragonal zirconia polycrystals

A. Domínguez-Rodríguez *, D. Gómez-García, M. Castillo-Rodríguez

*Departamento de Física de la Materia Condensada, Universidad de Sevilla, apartado 1065, 41080 Sevilla, Spain*

**Abstract**

In this paper, the origin of the threshold stress in pure 3 mol %-yttria tetragonal zirconia polycrystals (YTZP) is analysed in detail. At this regard, the two explanations reported in literature are discussed thoroughly. One of them invokes dislocation activity as the origin for such quantity, whereas the other one is based upon yttrium segregation at the grain boundaries.

Critical assessment for both of them is performed, and it has allowed concluding that the dislocation activity observed by Morita and Hiraga is just an artifact created during sample preparation. Thence, segregation at the grain boundaries seems to be the only mechanism accounting for a threshold stress and its temperature and grain size dependence.

**1. Introduction**

Since Wakai et al.1 showed the superplasticity in YTZP for the first time, a lot of effort has been devoted to this system; a good example of such intense research activity can be found in. 2–4 The steady-state strain rate $\dot{\varepsilon}$ at high temperature for super- plastically deformed ceramics is usually expressed by an equation of the form:

$$\dot{\varepsilon} = A\frac{Gb}{kT}\left(\frac{b}{d}\right)^{p}\left(\frac{\sigma}{G}\right)^{n}D \qquad (1)$$

where *A* is a dimensionless constant, *G* is the shear modulus, *b* is the magnitude of the Burgers vector, *k* is the Boltzmann's constant, *T* is the absolute temperature, *d* is the grain size, *p* is the inverse grain size exponent, *σ* is the stress, *n* is the stress exponent, and *D* is the appropriate diffusion coefficient given by: $D = D0 \exp(-Q/RT)$, where *D*0 is the pre-exponential factor which contains a frequency factor and *Q* is the activation energy of the mechanism controlling superplasticity, *R* is the gas constant. The parameters *n*, *p* and *Q* reported in the literature for YTZP have often shown different values for nominally identical mate- rials tested under similar experimental conditions.

Fig. 1 shows a plot of the strain rate versus stress, in arbitrary units, in which it is possible to observe the different domains of values for *n* and *Q*. At very low stresses, deformation occurs by Nabarro-Herring diffusional creep, with n values tending to 1 and *Q* around 450 kJ/mol.3,5 When the stress increases, the creep parameters become no longer constant: values of *n* between 3 and higher than 5 and values of *Q* as high as 700 kJ/mol are reported.3,4,6 At higher stresses and 450 kJ/mol respectively.3,4,6 To account for these different creep parameters, several explanations have been proposed. Among the different possible rate controlling mechanisms, three have been developed to account for the experimental creep results. (i) Two sequential processes: at high stresses, deformation occurs by grain boundary sliding (GBS) that is the slower process, whereas at low stresses, the deformation is controlled by an interface-reaction process.7,8 (ii) Interface-controlled diffusional creep; it is suggested that the flow process is controlled by a Coble-like mechanism provided the existence of grain-boundary dislocations, with two requisites: firstly, the core of those dislocations are perfect sources and sinks for vacancies and secondly, the dislocations are evenly spaced in the boundary planes so that all can climb at the same speed.9 (iii) GBS with a threshold stress; with the threshold stress, all the creep parameters can be explained by a single mechanism. The superplasticity is due to the GBS, being the accommodation processes controlled by the diffusion of point defects along the bulk.3,6 A detailed discussion of the different proposed theories can be found in.6

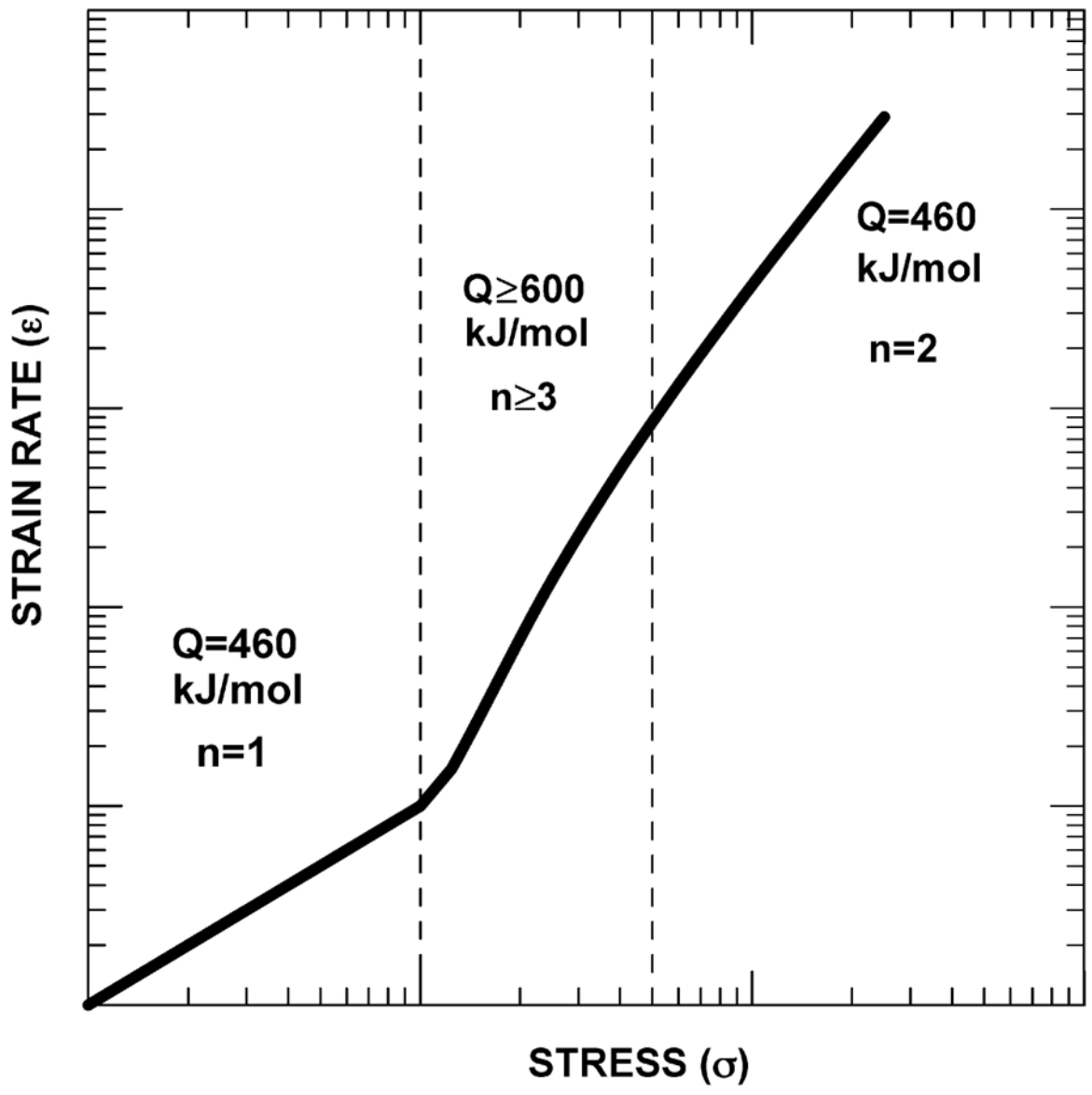

Fig. 1. Plot (in arbitrary units) of the strain rate vs. the stress to illustrate the different domains for plasticity behaviour reported in YTZP specimens.

Nowadays it is accepted that the most consistent explanation for superplasticity in YTZP is based upon grain boundary sliding with a threshold stress. The main objection to his explanation came from the lack of a physical model accounting for the ori- gin of this quantity and its peculiar grain size and temperature dependences. In fact, the equation describing the superplastic behaviour of YTZP is expressed as3,6 :

$$\dot{\varepsilon} = 3 \times 10^{10} \frac{(\sigma - \sigma_0)^2}{Td^2} \exp\left(-\frac{4.7\,\mathrm{eV}}{kT}\right) \qquad (2)$$

where $\sigma 0$ is a threshold stress, which has been determined experimentally as3,6:

$$\sigma_0 = \frac{3 \times 10^{-4}}{d} \exp\left(\frac{1.2\,\text{eV}}{kT}\right) \quad (3)$$

being $\sigma 0$ measured in MPa and *d* in µm.

In the paper, a critical analysis of its origin will be developed.

## 2. Experimental results and discussion

Two different models have been reported to this end. One of them is based upon the possible dislocation activity for plas- ticity accommodation, whereas the other one makes use of cation segregation as the source for a friction force for grain sliding. The second one is extensively reported in literature and it will be commented at the end briefly, whereas the first one is a more recent one and its assessment is the aim of this paper.

### *2.1. Intergranular dislocation*

In a recent paper by Morita and Hiraga10 these authors claim that *n* values as high as 5 and *Q* values equal to 680 kJ/mol, for stresses between 9 and 16 MPa, can be linked to the existence of a threshold stress for intragranular dislocation motion, being such motion the accommodation process for GBS. Such hypothesis is based on the observation of dislocation pile-ups as reported in.11

A dislocation pile-up is not a stable dislocation microstructure since strong dislocation repulsion of dislocation each other must occur. It remains stable as long as an external applied stress $\sigma$ exists. The shear stress $\tau$ at the head of the pile-up can be calculated using the equation11 :

$$\tau = \frac{Gb}{2L} N \quad (4)$$

where *L* is the pile-up length and *N* is the number of pile-up dislo- cations. Morita and Hiraga reported values of $\tau$ for the different pile-ups observed from 351 to 1260 MPa for nominal applied stresses from 15 to 50 MPa, Table 1 in.11 These values for the stress needed for pile-up stability are in full agreement with the reported values reported 400 MPa required for dislocation motion in yttria-tetragonal zirconia single crystals deformed at 1400 ◦C.12 Taking into account that for polycrystals the applied stress is 1/3 of the shear stress acting on the pile-up plane, the stress concentration ($\tau/\sigma$) lies in the range from 40 to 75 times the applied stress, too high to be explained by the stress concentration induced at multiple-grain junction during GBS. Several considerations must be done at this point: (i) As indicated by Morita and Hiraga, intragranular dislocations were frequently located around the multiple-grain junctions.11 Since stress demands for dislocation pile-ups generation are in between 350 and 1260 MPa, as observed by Morita and Hiraga, the internal stress in the superplastically deformed YTZP must reach values well over 350 MPa. This is more than one order of magnitude larger than the applied stress. As the applied stress is the sum of the internal stress and the effec- tive stress, it would mean that the effective stress becomes negative, which is clearly unphysical. (ii) Morita and Hiraga observed such pile-ups after the material had been cooled under load. This is consistent with the fact that Frank-Read sources can be generated under a stress field only. However, multipledislocation reactions take place when dislocation activity pro- ceeds. Thus, a complicated dislocation network is formed so that the internal stress field can be minimised. Since multiple reactions give rise to sessile segments, most dislocation seg- ments tend to be quite stable. No matter the cooling rate of the deformed sample, at least 70% of the dislocation density remains in the post-mortem microstructure. In consequence, the pile-up observed by Morita and Hiraga must be formed during cooling.

We have performed a traction experiment of YTZP specimens with a grain size equal to 0.38 μm (the grain size is measured using the mean intersect method with no numerical factor), at 1350 ◦C and 10−5 s−1 . These experimental conditions corre spond to the regime where the stress exponent is close to 3.13 To observe the microstructure created during deformation, the load was kept constant upon cooling. The cooling rate of the sample after deformation is displayed in Fig. 2. Two types of samples were prepared and observed through transmission electron microscopy (TEM).1 One of them corresponds to an as-received sample and the other one to the sample deformed and quenched according to the temperature profile displayed in Fig. 2. More than 500 grains have been characterized, changing the tilt-angle while observing under TEM to favour dislocation imaging in case they are present. Fig. 3 is a bright-field micrograph showing the typical microstructure of an as-received sample and Fig. 4 corresponds to the sample quenched after deformation. No dis- location activity was observed in any of the more than 500 grains studied.

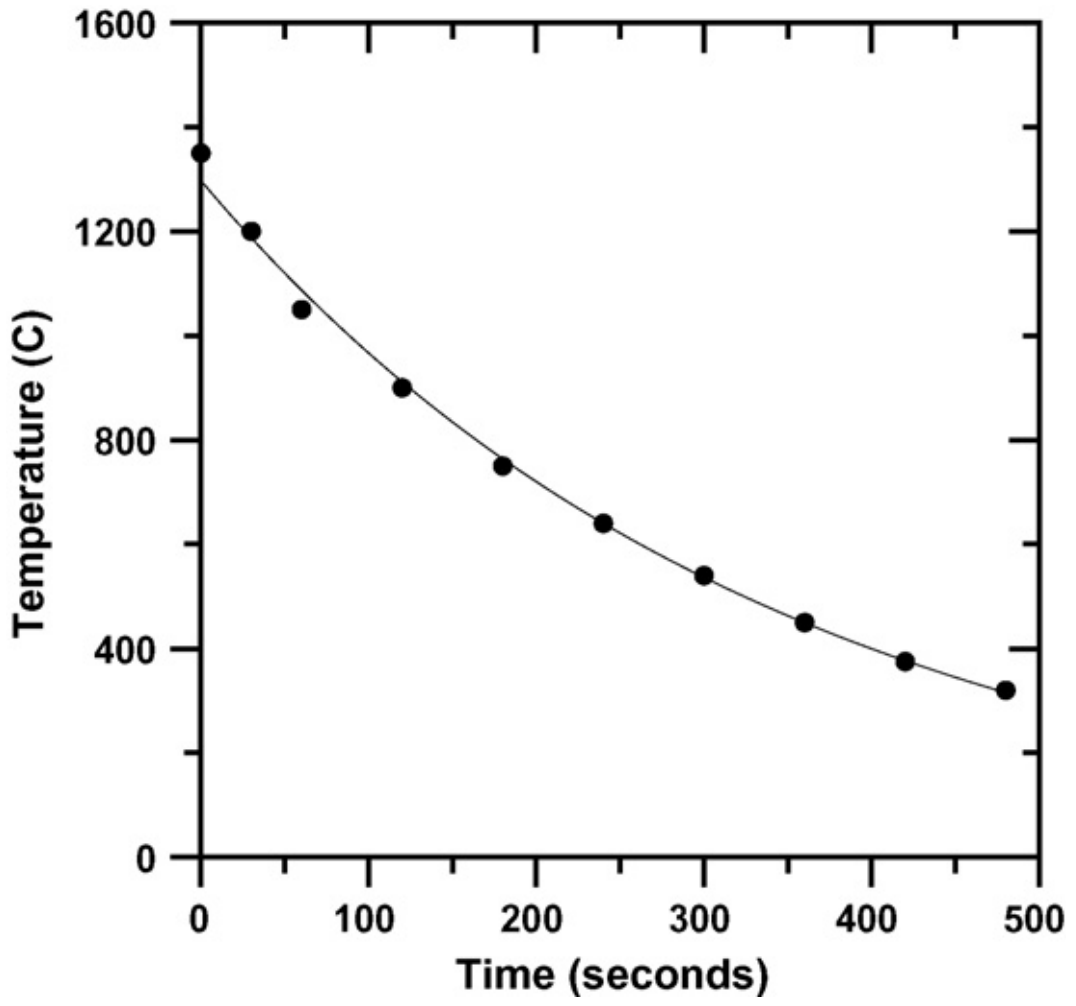

Fig. 2. Cooling rate of the traction test after deformation. As it can be observed, the temperature decreases from 1350 to 900 ◦ C in2 mn.

On the order hand, compression experiments have also been performed with two sets of superplastically deformed samples: one was cooled down slowly and the other one quenched and again no dislocations were observed in any case.

At this point, it should be remarked that the activity of dislocations cannot explain the grain size and temperature dependence of the threshold stress shown in equation (3). If dislocations were responsible for a threshold stress, the temperature dependence would be an Arrhenius-type one, since its activation is a thermally activated event. Clearly this is not the case, as displayed in Eq. (3).

All these considerations together with our TEM observations suggest us to say that the dislocation structure observed by Morita and Hiraga is an artifact.

### *2.2. Yttrium segregation at grain boundaries*

The alternative explanation for the existence of a threshold stress invokes the presence of cation segregation at the grain boundaries.

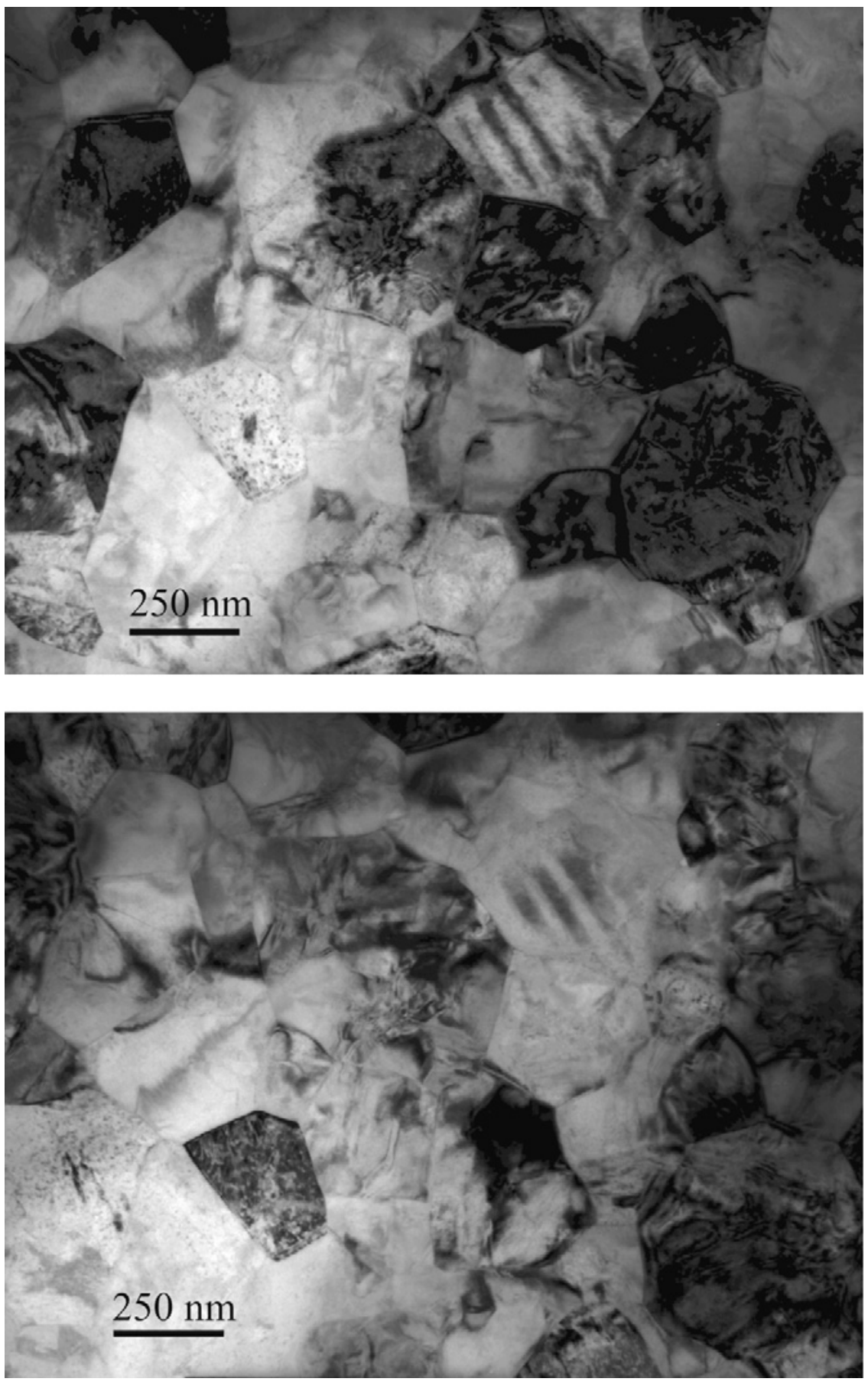

Fig. 3. Bright field transmission electron micrographs showing typical microstructure of the as-received sample. The two figures correspond to the same area with different tilt angles.

Nowadays it is well reported and accepted that yttrium segregates at grain boundaries or at the dislocation cores in YTZP. A thorough experimental analysis of the yttrium segregation at the grain boundaries is made by Boutz et al..14 The origin of this segregation is the relaxation of the elastic energy around yttrium atom as a consequence of the difference between the ionic radius of Y3+ and the Zr4+ , being around 20% bigger the Y3+ than Zr4+ one and has been proved experimentally by means of different techniques.14–20 The yttrium segregation depends on the temperature and bulk concentration19 and also on the grain size.15 Segregation of the cations to the grain boundaries changes the chemical composition as well as the electric space charge at the boundaries, consequently affecting the deformation processes in which grain boundary mobility is involved, such as grain boundary sliding. This, in turn, must affect the mechanical behaviour of the zirconia alloy, as shown recently in several papers. Details about how the seg- regation can influence the superplasticity in YTZP are found in references.21–23 A theoretical prediction for the temperature and grain size dependence is made, and a theoretical expression is achieved22,23 :

$$\sigma_0 = A\frac{1}{d}\exp\left(-\frac{2\Delta G}{kT}\right) \qquad (5)$$

Where *A* is a constant which depends on the details of the charge layer at the grain boundaries, and *ΔG* is the free enthalpy change for segregation from the bulk to the grain boundaries. Segregation is theremodynamically favoured when *ΔG* is neg- ative. This quantity depends on the shear modulus of the matrix and the segregating species and the misfit in ionic radius of the segregated and substituted species.

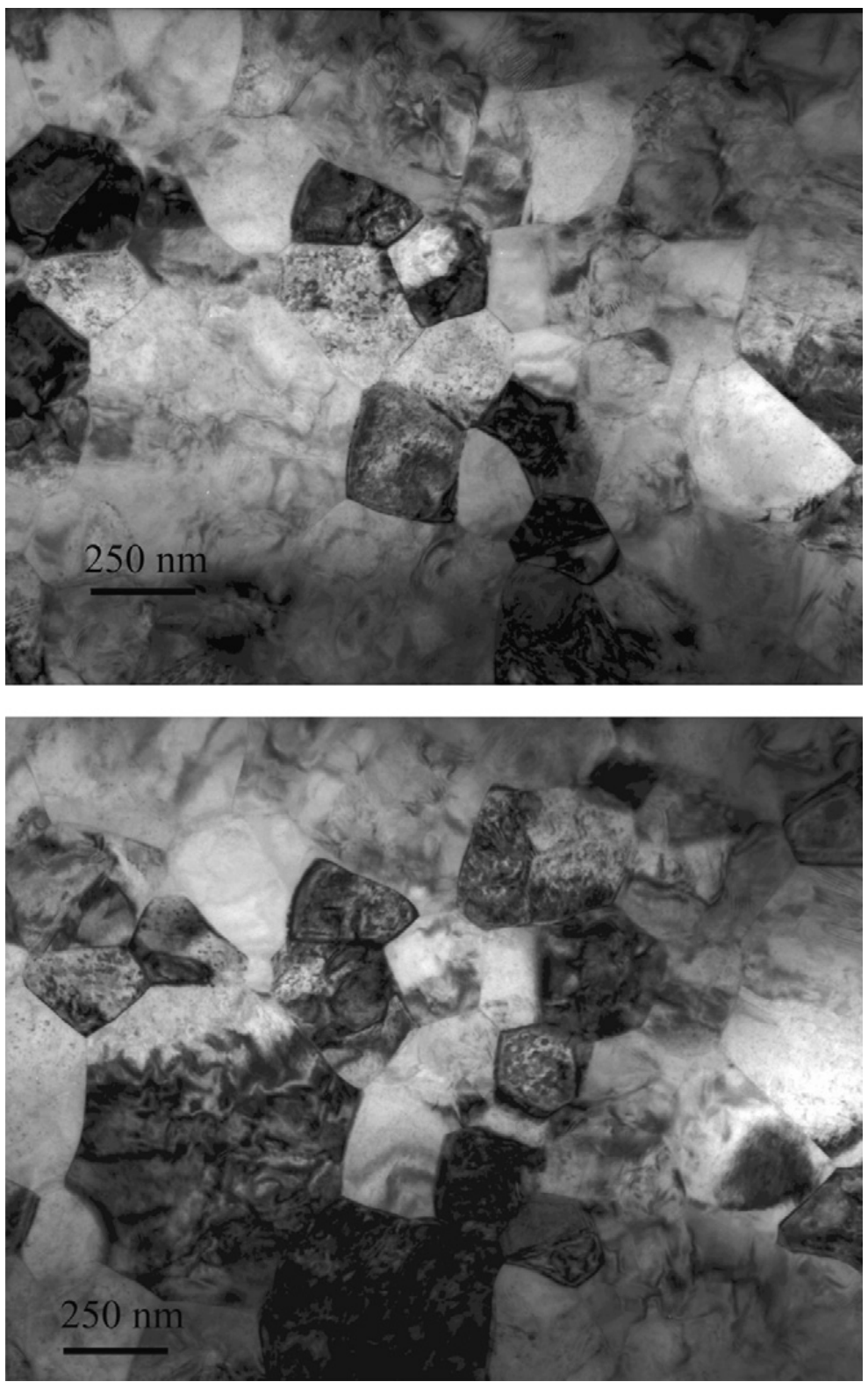

Fig. 4. Bright field transmission electron micrographs showing the typical microstructure of a sample deformed in traction and quenched at the cooling rate of the Fig. 2. The two figures correspond to the same tilted area in order to observe dislocations if present.

Although no experimental data are available for this mag- nitude, as far as we know, in the case of the solid solution of antimony in cooper, this energy is −65 kJ mol−1 .24 In this sys- tem, the difference between the atomic radius of Sb ($r1$ = 1.53 A˚) and that of cooper ($r$ = 1.23 A˚); the misfit between the two atoms is +0.194, very close to that found in yttia–zirconia solid solution.25 On the other hand, the shear modulus of both systems, 41 GPa for Sb–Cu,24 quite close to that of YTZP: 45 GPa.26 In consequence, it is logically accepted that the experimental value in equation (3) is a reasonable value. This value is in the cor- rect order of magnitude to that found in order systems, such as Y2 O3 –TiO2 , where Y3+ cations segregation has been put forward. The elastic energy is reported to be equal to −58 kJ/mol at room temperature.27 These values are fully consistent with the YTZP system, in which ionic radius are very close to those for Y2 O3 –TiO2. On the other hand, Yoshida et al.28 have deformed nanocrystalline monoclinic zirconia with grain size equal to 65 nm, in the range of temperatures between 1273 and 1373 K and although the phase in that range of temperatures is tetragonal, there is no yttrium content in them. The main results of this work is that the stress exponent is 2.5 in all range of stresses used in this work and no threshold stress exists, in agreement with the assumption that this threshold stress is due to the yttrium segregation at the grain boundaries.

As it can be observed in this Eq. (5), the impurities segregation at the grain boundaries can not only justify the existence of the threshold stress but also give the grain size and temperature dependence found experimentally.

## 3. Conclusions

A critical assessment of the different reported models to account for the threshold stress in YTZP is carried out. The experimental tests to validate the dislocation-driven model show that this explanation is not consistent with the experimental facts; i.e. no dislocation activity is detected, and anyway, it cannot explain the temperature and grain size dependence of the creep equation.

The cation segregation model can explain such dependencies and it is in full agreement with the fact that no threshold stress is measured in monoclinic zirconia, in which no cation segregation occurs.

## Acknowledgements

This work was supported by the Spanish grants MAT2003- 04199-C02-02 and MAT2006-10249-C02-02 (Ministerio de Educación y Ciencia, Spain). Spanish-Japanese cooperation grant sponsored by the Spanish CSIC and the Japan Society for the Promotion of Science (2004JP0119) is acknowledged. The technical support provided by the research team of Prof. Wakai's group during the stay of two authors in Tokyo Institute of Technology is gratefully acknowledged, as well as Professor Wakai for encouraging discussion.

## References

1. Wakai, F., Sakaguchi, S. and Matsuno, Y., Superplasticity of Yttria- stabilized tetragonal ZrO2 polycrystals. *Adv. Ceram. Mater*, 1986, **1**, 259–263.

2. Chokshi, A. H., Mukherjee, A. K. and Langdon, T. G., Superplasticity in advanced materials. *Mater. Sci. Eng. R.*, 1993, **10**, 237–274.

3. Jimenez-Melendo, M., Domiınguez-Rodrıiguez, A. and Bravo-Leon, A., Superplastic flow in fine-grained yttria-stabilized zirconia polycrystals: con stitutive equation and deformation mechanisms. *J. Am. Ceram. Soc.*, 1998, **81**, 2761–2776.

4. Domiınguez-Rodrıiguez, A., Bravo-Leo´ n, A., Ye, J. D. and Jimenez-Melendo, M., Grain size and temperature dependence of the threshold stress for super- plastic deformation in yttria-stabilized zirconia polycrystals. *Mater. Sci. Eng.*, 1998, **A247**, 97–101.

5. Bravo-Leo´ n, A., Jimenez-Melendo, M. and Dominguez-Rodriguez, A., High temperature plastic deformation at very low stresses of fine-grained Y2 O3 -partially stabilized ZrO2 . *Scripta Mater.*, 1996, **35**, 551–555.

6. Jimenez-Melendo, M. and Dominguez-Rodriguez, A., High temperature mechanical characteristics of superplastic Yttria stabilized Zirconia. An examination of the flow process. *Acta Mater.*, 2000, **48**, 3201–3210.

7. Boutz, M. M. R., Winnubst, A. J. A., Burgraaf, A. J., Nauer, M. and Carry, C., Low temperature superplastic flow of yttria stabilised zirconia polycrystals. *J. European Ceram. Soc.*, 1994, **13**, 103–111.

8. Owen, D. M. and Chokshi, A. H., The high temperature mechanical char- acteristics of superplastic 3 mol % Yttria stabilized Zirconia. *Acta Mater.*, 1998, **46**, 667–679.

9. Berbon, M. Z. and Langdon, T. G., An examination of the flow process in superplastic Yttria-stabilized tetragonal Zirconia. *Acta Mater.*, 1999, **47**,2485–2495.

10. Morita, K. and Hiraga, K., Critical assessment of high-temperature defor- mation and deformed microstructure in high-purity tetragonal zirconia containing 3 mol % yttria. *Acta Mater.*, 2002, **50**, 1075–1085.

11. Morita, K. and Hiraga, K., Deformed substructures in fine-grained tetragonal zirconia. *Phil. Mag. Lett.*, 2001, **81**, 311–319.

12. Munoz, A., Gomez-Garcia, D., Dominguez-Rodriguez, A. and Wakai, F., High temperature plastic anisotropy of YTZ-single crystals. *J. Eur. Ceram. Soc.*, 2002, **22**, 2609–2613.

13. Zapata-Solvas, E., Gomez-Garcia, D., Garcia-Ganan, C. and Dominguez Rodriguez, A., High temperature creep behaviour of 4 mol % yttria tetragonal zirconia polycrystals (4-YTZP) with grain sizes within 0.38 < *d* < 1.15 f.Lm. *J. Eur. Ceram. Soc.*, 2007, **27**, 3325–3329.

14. Boutz, M. M. R., Chen, C. S., Winnubst, A. J. A. and Burggraaf, A. J., Characterization of grain boundaries in superplastically deformed Y-TZP ceramics. *J. Am. Ceram. Soc.*, 1994, **77**, 2632–2640.

15. Theunissen, G. S. A. M., Winnubst, A. J. A. and Burggraaf, A. J., Surface and grain boundary analysis of doped zirconia ceramics studied by AES and XPS. *J. Mater. Sci.*, 1992, **27**(18), 5057–5066.